\documentclass{aa}
\usepackage{graphicx}
\usepackage{natbib}
\usepackage{txfonts} 
\begin{document}

\title{
Discovery of the Binary Pulsar PSR\,B1259$-$63 in Very-High-Energy
Gamma Rays around Periastron with H.E.S.S.}

\author{F. Aharonian\inst{1}
 \and A.G.~Akhperjanian \inst{2}
 \and K.-M.~Aye \inst{3}
 \and A.R.~Bazer-Bachi \inst{4}
 \and M.~Beilicke \inst{5}
 \and W.~Benbow \inst{1}
 \and D.~Berge \inst{1}
 \and P.~Berghaus \inst{6}\thanks{Universit\'e Libre de 
 Bruxelles, Facult\'e des Sciences, Campus de la Plaine, CP230, Boulevard
 du Triomphe, 1050 Bruxelles, Belgium}
 \and K.~Bernl\"ohr \inst{1,7}
 \and C.~Boisson \inst{8}
% ...
}

\institute{
Max-Planck-Institut f\"ur Kernphysik, Heidelberg, Germany
\and
Yerevan Physics Institute, Yerevan, Armenia
\and
University of Durham, Department of Physics, Durham, U.K.
\and
Centre d'Etude Spatiale des Rayonnements, CNRS/UPS, Toulouse, France
\and
Universit\"at Hamburg, Institut f\"ur Experimentalphysik, Hamburg,
Germany
\and
APC, Paris, France\thanks{UMR 7164 (CNRS, Universit\'e Paris VII, CEA,
Observatoire de Paris)}
\and
Institut f\"ur Physik, Humboldt-Universit\"at zu Berlin, Berlin,
Germany
\and
LUTH, UMR 8102 du CNRS, Observatoire de Paris, Meudon, France
% ...
}

\date{Accepted 2 June 2005}

\offprints{Stefan Schlenker,\\
\email{schlenk@physik.hu-berlin.de}}

\abstract{
We report the discovery of very-high-energy (VHE) $\gamma$-ray
emission of the binary system PSR\,B1259$-$63\,/\,SS\,2883 of a radio
pulsar orbiting a massive, luminous Be star in a highly eccentric
orbit. The observations around the 2004 periastron passage of the
pulsar were performed with the four 13\,m Cherenkov telescopes of the
H.E.S.S.\ experiment, recently installed in Namibia and in full
operation since December 2003. Between February and June 2004, a
$\gamma$-ray signal from the binary system was detected with a total
significance above $13\,\sigma$. The flux was found to vary
significantly on timescales of days which makes PSR\,B1259$-$63 the
first variable galactic source of VHE $\gamma$-rays observed so
far. Strong emission signals were observed in pre- and post-periastron
phases with a flux minimum around periastron, followed by a gradual
flux decrease in the months after. The measured time-averaged energy
spectrum above a mean threshold energy of 380\,GeV can be fitted by a
simple power law $F_0(E/1\,\rm TeV)^{-\Gamma}$ with a photon index
$\Gamma = 2.7\pm0.2_\mathrm{stat}\pm0.2_\mathrm{sys}$ and flux
normalisation $F_0 = (1.3 \pm 0.1_\mathrm{stat} \pm 0.3_\mathrm{sys})
\times 10^{-12}\,\rm TeV^{-1}\,\rm cm^{-2}\,\rm s^{-1}$. This
detection of VHE $\gamma$-rays provides unambiguous evidence for
particle acceleration to multi-TeV energies in the binary system. In
combination with coeval observations of the X-ray synchrotron emission
by the {\em RXTE} and {\em INTEGRAL} instruments, and assuming the VHE
$\gamma$-ray emission to be produced by the inverse Compton mechanism,
the magnetic field strength can be directly estimated to be of the
order of 1\,G.

}

\authorrunning{F. Aharonian et al.}
\titlerunning{
Discovery of PSR B1259$-$63 in VHE $\gamma$-rays around Periastron}

\maketitle

\section{Introduction}

PSR\,B1259$-$63\,/\,SS\,2883 is a binary system consisting of a
$\sim$48\,ms pulsar in orbit around a massive B2e companion star
\citep{Johnston:1,Johnston:2}. The highly eccentric orbit of the
pulsar places it just $\sim 10^{13} \rm cm$ from the companion during
periastron every $\sim$3.4\,years. Be stars are known to have
non-isotropic stellar winds forming an equatorial disk with enhanced
mass outflow \citep[e.g.][]{Waters:1}. In the case of PSR\,B1259$-$63,
timing measurements suggest that the disk is inclined with respect to
the orbital plane \citep{Wex:1}, probably because the neutron star
received a substantial birth kick, causing the pulsar to cross the
disk two times near periastron. These unique properties make the
binary system PSR\,B1259$-$63 an excellent laboratory for the study of
pulsar winds interacting with a changing environment in the presence
of an extremely intense photon field. The synchrotron origin of
optically thin unpulsed radio emission detected from this source, in
particular during the periastron passage
\citep[e.g.][]{Johnston:3,Connors:1}, indicates acceleration of
electrons to relativistic energies. The acceleration process has been
argued to be most efficient when the pulsar passes through the
equatorial disk \citep{Ball:3}. The basic features of such a system
(``binary plerion'') in the context of higher energy X- and
$\gamma$-radiation components have been comprehensively discussed by
\citet{Tavani:2}.

The intense photon field provided by the companion star not only plays
an important role in the cooling of relativistic electrons but also
serves as perfect target for the production of high energy
$\gamma$-rays through inverse Compton (IC) scattering
\citep{Tavani:1,Kirk:1,Ball:1,Ball:2,Murata:1}. Some of these
emission models predict wind powered shock acceleration of electrons
to multi-TeV energies, radiating predominantly through the
synchrotron and IC channels, with the main energy release in the X-
and high energy $\gamma$-ray bands, respectively.

The unpulsed non-thermal X-ray emission detected from PSR\,B1259$-$63
throughout its orbital phase in 1992 to 1996 by the {\em ROSAT} and
{\em ASCA} satellites \citep{Cominsky:1,Kaspi:1,Hirayama:1} generally
supports the synchrotron origin of X-rays. The spectrum of the
synchrotron radiation seems to extend to hard X-rays/low energy
$\gamma$-rays as shown by {\em OSSE} \citep{Grove:1} and recently
confirmed by observations with the {\em INTEGRAL} satellite
\citep{Shaw:1}.

Since the companion star provides the dominant source of photons for
IC scattering, the target photon density is well known throughout the
entire orbit. Therefore, the ratio of X-ray flux to high energy
$\gamma$-ray flux depends only on the strength of the ambient magnetic
field. Although the latter can be estimated within a general
magneto-hydrodynamic treatment of the problem, it contains large
uncertainties which affect the estimate of the IC $\gamma$-ray flux as
$F_\gamma \propto B^{-2}$.

\citet{Kirk:1} studied the light curves of very-high-energy (VHE)
$\gamma$-rays under the assumption of a $1/r$ dependence of the
magnetic field which implies that the ratio of the energy density of
the photon field to that in the magnetic field $B$ is independent of
orbital phase. They also assumed that the position of the termination
shock as well as the strength of the magnetic field is not affected by
the disk of the B2e star. Under such assumptions, they predicted an
asymmetric $\gamma$-ray light curve with respect to periastron
(because of the inclination of the orbit with respect to the line of
sight and the dependence of the inverse Compton $\gamma$-ray
emissivity on the scattering angle), with an increase towards
periastron and monotonic decrease after the passage of
periastron. However, one might possibly expect significant deviation
from such a simplified picture given the apparent strong impact of the
disk on the pulsar wind termination as seen in the X-ray light curve
\citep{Tavani:2}. Moreover, during the time periods of interaction of
the pulsar wind with the equatorial disk one may expect, in addition
to the IC $\gamma$-rays, a new component of $\gamma$-radiation
associated with interactions of accelerated electrons and possibly
also protons with the dense ambient gas \citep{Kawachi:1}. Up to now,
the theoretical understanding of the properties of this complex
system, involving pulsar and stellar winds interacting with each
other, is quite limited because of the lack of constraining
observations.

Nevertheless, the fortunate combination of: (1) the high spin-down
luminosity of the pulsar, $L=8.3\times 10^{35}\,\rm erg/s$, which is
partially converted into populations of ultra-relativistic particles,
(2) the presence of the intense target photon field provided by the
companion star with energy density $\approx\,0.9\,(R(t)/10^{13}\,\rm
cm)^{-2}\,\rm erg/cm^3$ (where $R(t)$ is the spatial separation
between the pulsar and the companion star), and (3) the relatively
small distance to the source ($d \approx 1.5\,\rm kpc$) makes this
object a very attractive candidate for VHE $\gamma$-ray emission.

Previous observations of PSR\,B1259$-$63 in VHE $\gamma$-rays, apart
from its periastron passage were performed using the {\em CANGAROO I}
and {\em CANGAROO II} detectors, but did not result in significant
signals and provided upper limits at $13\%$ of the flux from the Crab
Nebula \citep[see][ and references therein]{Kawachi:1}. The first
significant detection in VHE $\gamma$-rays based on preliminary
analysis results was reported in \citet{IAU} a few days prior to the
periastron passage in February 2004 by the High Energy Stereoscopic
System (H.E.S.S.) to allow for target of opportunity observations of
other instruments. The analysis of the data from this initial
detection and the data obtained in the subsequent H.E.S.S.\
observation campaign on PSR\,B1259$-$63 is presented in this paper.

\section{VHE $\gamma$-ray Observations and Results}

\subsection{Observations and Analysis}

The observations from February to June 2004 were performed with the
High Energy Stereoscopic System (H.E.S.S.), consisting of four imaging
atmospheric Cherenkov telescopes \citep{status} located in Namibia, at
$23\degr 16'$~S $16\degr 30'$~E in 1800\,m above sea level. Each
telescope has a tesselated spherical mirror with 13\,m diameter and
$107\,\rm m^2$ area \citep{optics1, optics2} and is equipped with a
camera of 960 $0.16\degr$-photomultiplier tubes providing a total
field of view of $5\degr$ in diameter \citep{camera}. During the
stereoscopic observations, an array trigger requires the simultaneous
detection of air-showers by several telescopes at the hardware level,
allowing a suppression of background events \citep{trigger}.

All observations were carried out in moonless nights tracking sky
positions with an alternating offset of typically $\pm0.5^\circ$ in
declination relative to the source (the {\em wobble} mode) in time
intervals of 28\,minutes duration. This allows to determine the
background from the same field of view and one can omit off-source
observations, effectively doubling the observation time. Due to the
serendipitous discovery of another source in the field of view around
PSR\,B1259$-$63 \citep[see][]{HESS1303}, for all observations
subsequent to the 14th of May 2004 (MJD 53139) the array pointing was
changed to a position $\sim0.6$\degr\ north of PSR\,B1259$-$63 and an
alternating wobble offset of $0.5$\degr\ in right ascension (instead
of declination) was used.

The data set, selected on the basis of standard quality criteria, has
a dead time corrected exposure (live time) of 48.6\,h and a mean
zenith angle of 42.7$\degr$. The corresponding mean threshold energy
defined by the peak $\gamma$-ray detection rate for a source with a
Crab-like spectrum \citep{hegra_crab:1} after selection cuts was
estimated to be 380\,GeV. In the phase prior to the periastron passage
(7.8\,h live time), due to technical problems, data from only three
telescopes were considered, and for the post-periastron phase (41.9\,h
live time) data from the full telescope array were used.

After the calibration of the recorded air shower data
\citep{calibration}, each telescope image was parametrised by its
centre of gravity and second moments \citep{Hillas} followed by the
stereoscopic reconstruction of the shower geometry providing an
angular resolution of $\sim0.1\degr$ for individual $\gamma$-rays.
The $\gamma$-ray energy was estimated from the image intensity and the
shower geometry with a typical resolution of $\sim15\%$. In order to
reject the vast background of cosmic-ray showers, $\gamma$-ray
candidates are selected using cuts on image shape scaled with their
expectation values obtained from Monte Carlo simulations. The cuts
used for this analysis were optimised on simulations of a $\gamma$-ray
point source with a flux level of 10\% of the Crab Nebula VHE
$\gamma$-ray flux, allowing 41.2\% of the $\gamma$-rays to be retained
while rejecting more than 99.9\% of the cosmic-ray air showers. A more
detailed description of the analysis techniques can be found in
\citet{PKS2155}.

\subsection{Detection}

\begin{figure}[t]
\centering
\resizebox{\hsize}{!}{\includegraphics{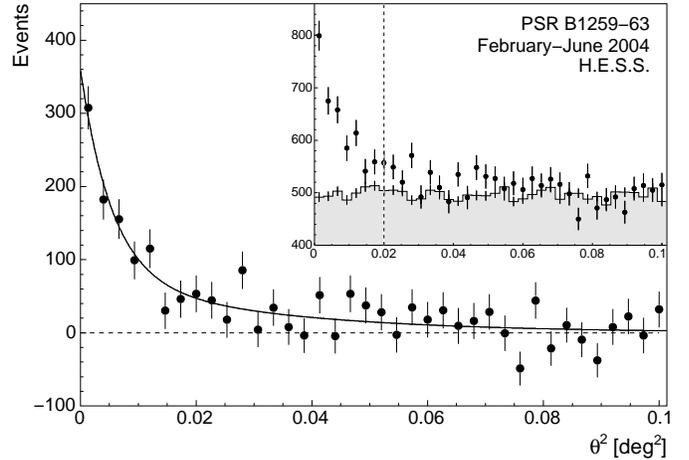}}
\caption{
Main figure: Distribution of background-subtracted $\gamma$-ray
candidates in $\theta^2$ where $\theta$ is the angular distance
between the reconstructed shower direction and the pulsar position for
the complete 2004 data set. The solid line indicates the distribution
expected for a point source of $\gamma$-rays. Inset: Distributions in
$\theta^2$ for ``on-source'' $\gamma$-ray candidates relative to the
source position (full points) and for all background control regions
(histogram, scaled with the background normalisation factor
$\alpha$). An uniform background results in a flat distribution in
$\theta^2$. The dashed vertical line indicates the angular cut applied
($\theta^2 < \theta^2_{\rm cut} = 0.02\,\rm deg^2$).}
\label{fig_thetasquared}
\end{figure}

Figure \ref{fig_thetasquared} shows the distribution of the squared
angular distance $\theta^2$ of excess events relative to the position
of PSR\,B1259$-$63 for the whole data set. The circle with a radius
corresponding to the angular cut
$\theta^2_\mathrm{cut}=0.02\,\mathrm{deg}^2$ in the field of view
around the source position was considered as the {\em on-region}. The
background was estimated from several non-overlapping circles of the
same radius ({\em off-regions}) with the same angular distance from
the camera centre, allowing corrections due to the varying camera
acceptance to be omitted \citep{hegra_CasA}. The clear excess in the
direction of the pulsar has a significance of 13.8\,$\sigma$ and is
consistent with a distribution obtained from a simulated $\gamma$-ray
point source. The $\gamma$-ray signal from the direction of
PSR\,B1259$-$63 was detected in most of the darkness periods from
February to June 2004, for which the results are summarised in
Table~\ref{table_periods}.

\begin{table}[b]
\caption{\label{table_periods}
Results of H.E.S.S.\ observations on PSR\,B1259$-$63. For each
darkness period the number of telescopes used in the analysis
$N_{\mathrm{tel}}$, the live time $t_{\mathrm{live}}$, the
significance $S$ calculated according to \citet{LiMa}, the number of
counts within the on-source ($N_{\mathrm{on}}$) and off-source
($N_{\mathrm{off}}$) region(s), the background normalisation $\alpha$,
and the number of detected $\gamma$-rays ($N_{\mathrm{\gamma}}$) are
listed. The background normalisation $\alpha$ was determined by the
number of off-regions, which do not coincide with other $\gamma$-ray
sources in the field of view. Therefore $\alpha$ is not the same for
all data subsets because it depends on the distribution of wobble
offsets used when obtaining the data.}
\centering
\resizebox{\hsize}{!}{\begin{tabular}{crrrrrrr}
\hline\hline
Period & $N_{\mathrm{tel}}$ & $t_{\mathrm{live}}$ & $S$ & $N_{\mathrm{on}}$ & $N_{\mathrm{off}}$ & $\alpha$ & $N_{\mathrm{\gamma}}$\\
2004 & & [h] & [$\mathrm{\sigma}$] & & & &\\
\hline
February & 3 & 7.9  & 9.8  & 691    & 2\,591  & 0.172 & 246$\pm$25 \\
March    & 4 & 17.8 & 7.2  & 1\,740 & 8\,459  & 0.169 & 307$\pm$43 \\
April    & 4 & 5.1  & 7.8  & 644    & 2\,705  & 0.167 & 193$\pm$25 \\
May      & 4 & 9.9  & 5.2  & 910    & 4\,499  & 0.167 & 160$\pm$31 \\
June     & 4 & 9.1  & 1.8  & 722    & 4\,027  & 0.167 & 51$\pm$28 \\
\hline
Total    &   & 49.8 & 13.8 & 4\,707 & 22\,281 & 0.168 & 955$\pm$69\\
\hline
\end{tabular}}
\end{table}

\begin{figure}[t]
\centering
\resizebox{0.99\hsize}{!}{\includegraphics{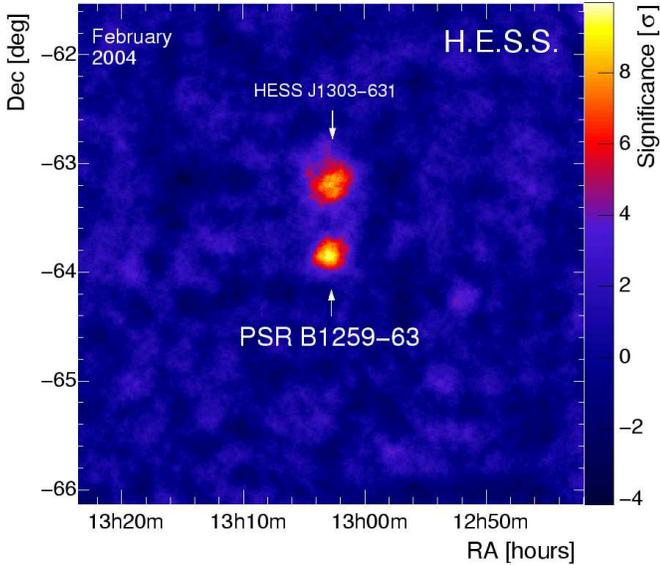}}
\caption{
Significance sky-map centred on the position of PSR\,B1259$-$63 for
the H.E.S.S.\ February data. The excess $\sim 0.6\degr$ north of the
pulsar is the newly discovered unidentified TeV source
HESS\,J1303$-$631 \citep{HESS1303}. The bins are correlated within a
circle of radius $\theta_\mathrm{cut} = 0.02\,\rm deg^2$. The
background for each position in the field of view was estimated from a
ring around this position correcting for the radial acceptance of the
instrument within the field of view and background contamination from
the two $\gamma$-ray sources.}
\label{fig_skymap}
\end{figure}

A 2D-analysis of the H.E.S.S.\ field of view around PSR\,B1259$-$63
was performed using the ring background technique as alternative
background estimation method. In this method, for each bin, the number
of on-events $N_\mathrm{on}$ was derived by integrating the bin
content within a circle of radius $\theta_\mathrm{cut}$
(on-region). The number of background events $N_\mathrm{off}$ was
estimated from a ring around the bin position with mean radius
$0.6\degr$ and an area 7 times larger than the area of the
on-region. The normalisation $\alpha$ was corrected for the decrease
of the radial acceptance of the cameras towards the edge of the field
of view. Figure~\ref{fig_skymap} shows the significance sky-map of a
$2.3\degr$ field of view around PSR\,B1259$-$63 for the February
data. The excess at the pulsar position
($N_\mathrm{\gamma}=N_\mathrm{on}-\alpha N_\mathrm{off}=227\pm27$,
with a significance $S=9.1\,\sigma$) is consistent with the excess
given in Table~\ref{table_periods}. The additional broad excess $\sim
0.6\degr$ north of the pulsar is the unidentified TeV source
HESS\,J1303$-$631 discovered in the same field of view
\citep{HESS1303}. The resulting bias in the background estimation due
to both sources was corrected in the analysis by excluding events from
a circle with radius $0.4\degr$ around each source from the background
estimation.

In order to derive the position of the pulsar excess and to check for
possible source extension, the data was reanalysed using hard cuts
requiring a minimal camera image intensity of 200\,photo-electrons
which significantly improves the angular resolution and drastically
reduces the cosmic ray background at the expense of a higher energy
threshold of 750\,GeV. The uncorrelated two dimensional excess
distribution for the whole data set was fitted assuming a radially
symmetric, Gaussian source intensity profile
\[
I(\theta) \propto e^{-\theta^2/2 \sigma_\mathrm{s}^2}
\]
convolved with the point spread function of the detector. A possible
bias from the second source within the field of view was corrected for
by including an additional non-symmetric two dimensional Gaussian
excess distribution in the fit, centred at the position of
HESS\,J1303$-$631 and allowing the orientation and width to vary
freely. The resulting pulsar excess position of $\rm
RA\,13^\mathrm{h}2^\mathrm{m}49\fs3 \pm 2\fs3_\mathrm{stat}$, $\rm
Dec\,-63\degr49\arcmin53\arcsec \pm 17\arcsec_{\rm stat}$ is
consistent with the position of PSR\,B1259$-$63 ($\Delta\mathrm{RA} =
9\arcsec \pm 15\arcsec_{\rm stat}$, $\Delta\mathrm{Dec} = 16\arcsec
\pm 17\arcsec_{\rm stat}$) within statistical errors. The systematic
uncertainty on the absolute pointing of the telescopes has been
estimated in \citet{pointing} to be $\sim20\arcsec$. A limit for the
source extension $\sigma_\mathrm{s}$ was found to be $<33\arcsec$ at
$95\%$ confidence level which corresponds to 0.24\,pc at an assumed
distance of 1.5\,kpc. Note that a possible source confusion with the
variable hard X-ray source 1RXP\,J130159.6$-$635806 \citep{Kaspi:1},
located $9\arcmin$ southeast from the pulsar, can be firmly excluded.

\subsection{Energy Spectra}

\begin{figure}[p]
\centering
\resizebox{\hsize}{!}{\includegraphics{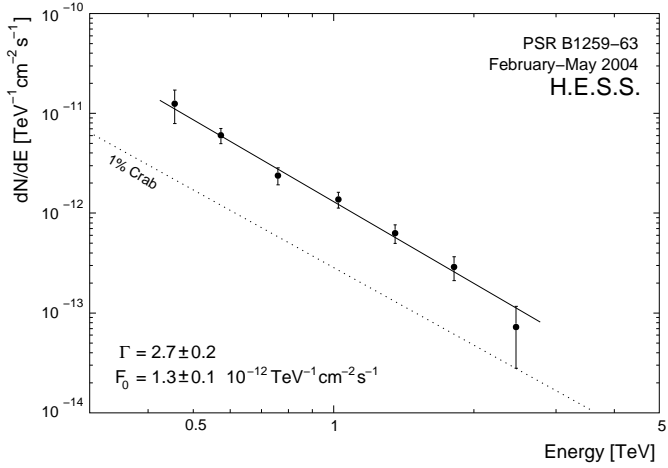}}
\caption{
Differential energy spectrum $\mathrm{d}N/\mathrm{d}E$ of
$\gamma$-rays from PSR\,B1259$-$63 using H.E.S.S.\ 2004 data from the
periods with significant detection of the pulsar (February--May). The
solid line shows the power-law fit to the spectrum (see also
Table~\ref{table_spectrum}).}
\label{fig_spectrum}
\end{figure}

\begin{figure}[p]
\centering
\resizebox{0.885\hsize}{!}{\includegraphics{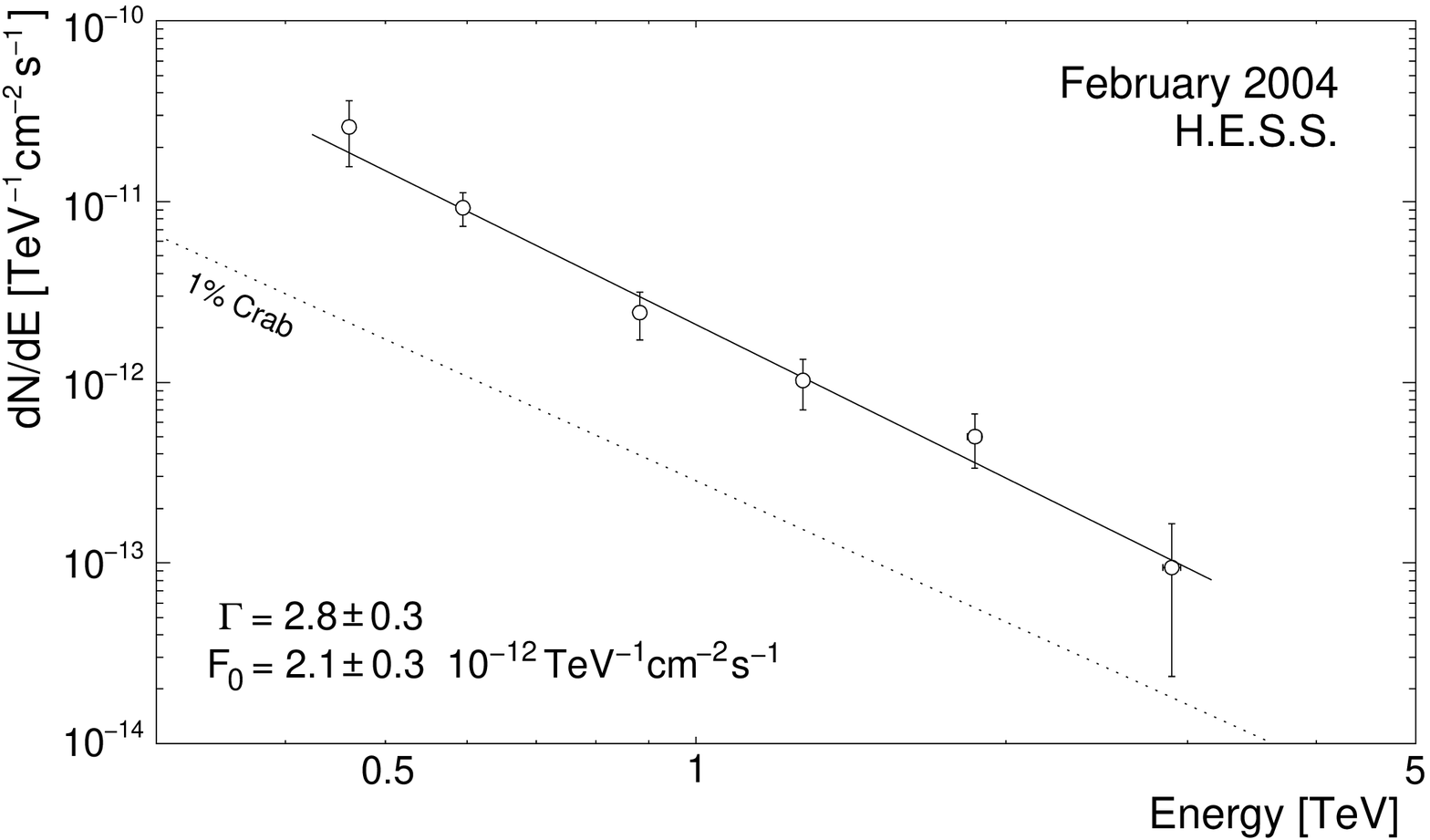}}
\resizebox{0.885\hsize}{!}{\includegraphics{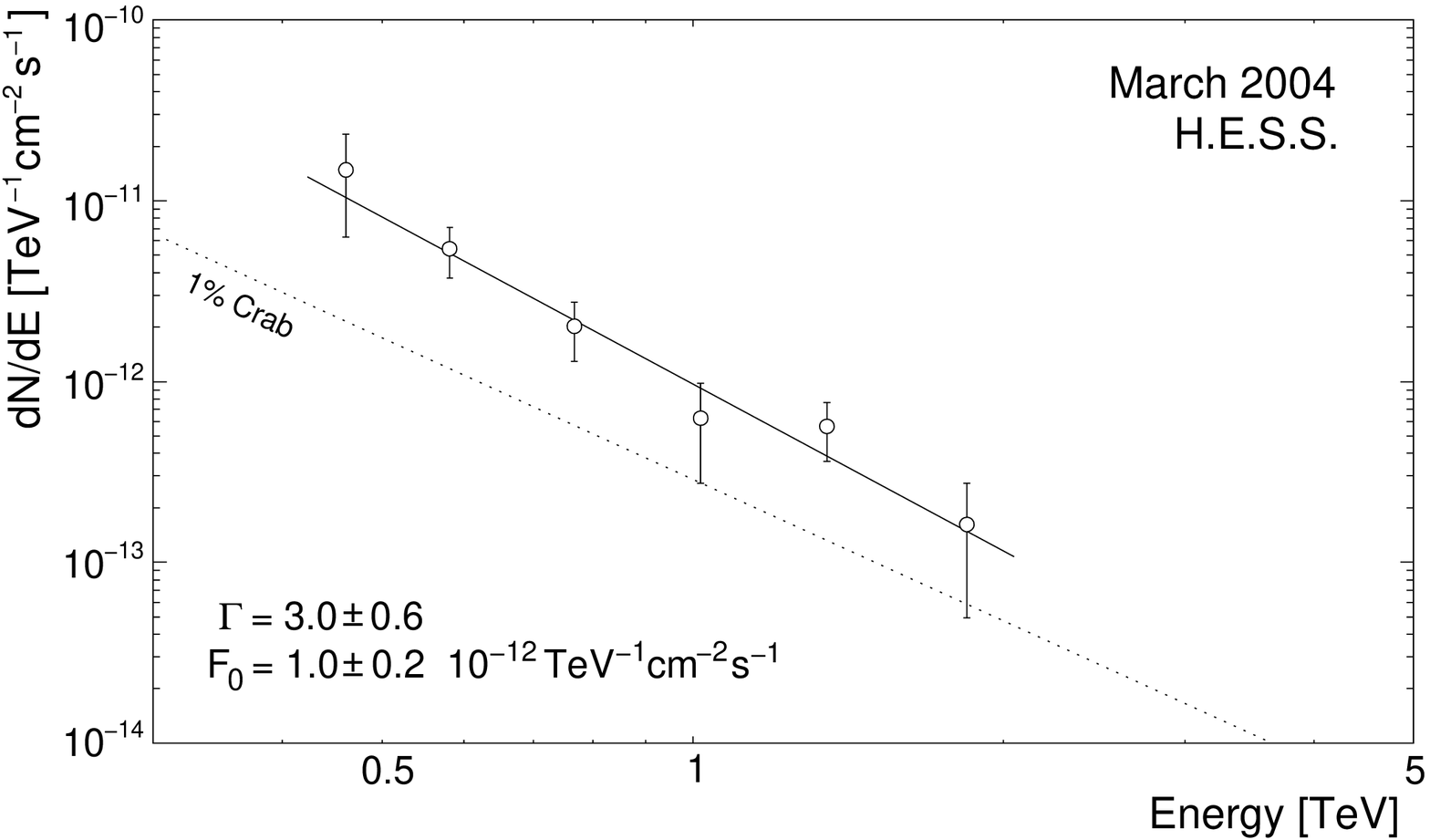}}
\resizebox{0.885\hsize}{!}{\includegraphics{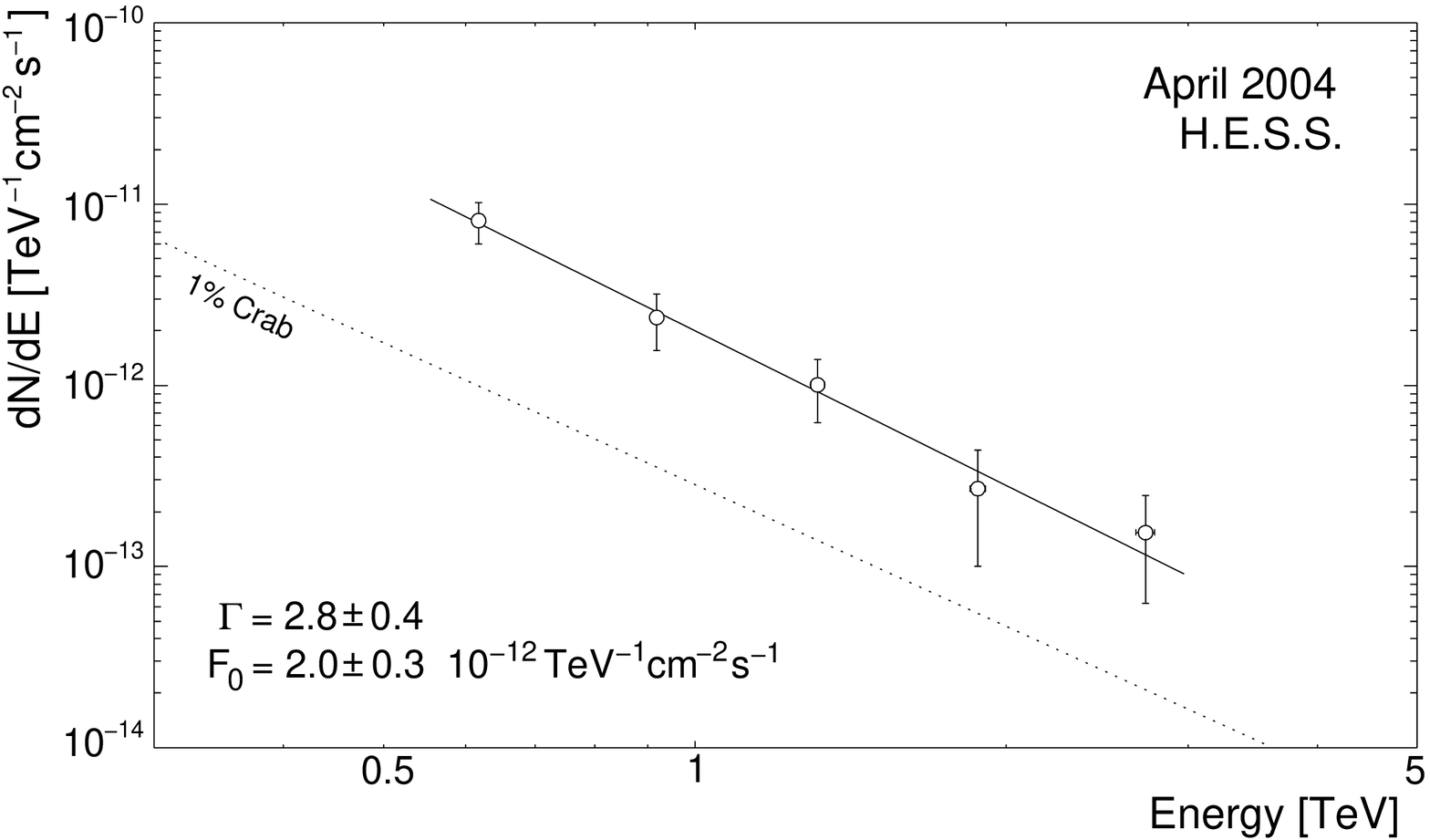}}
\caption{
Differential energy spectra of $\gamma$-rays from PSR\,B1259$-$63 for
data of the periods February, March and April (from top to
bottom). The straight line in each plot indicates the power-law fit to
the particular spectrum for which the results are listed in Table
\ref{table_spectrum}. Note, that the photon index $\Gamma$ is constant
within the statistical errors.}
\label{fig_spectra}
\end{figure}

The measured differential energy spectrum derived from all darkness
periods with a significant detection of the source (February to May)
is shown in Fig.~\ref{fig_spectrum}. A power-law fit
\begin{equation}\label{powerlaw}
F(E)=\mathrm{d}N/\mathrm{d}E =
F_0\left(\frac{E}{\mathrm{1\,TeV}}\right)^{-\Gamma}
\end{equation}
of this spectrum yields a photon index $\Gamma=2.7\pm0.2_{\rm stat}$
and $F_0 = (1.3\pm0.1_{\rm stat})\times 10^{-12}\rm cm^{-2}\rm
s^{-1}\rm TeV^{-1}$. The integral flux above the mean threshold energy
is $F(E>380\,\rm GeV) = (4.0\pm0.4_{\rm stat})\times 10^{-12}\rm
cm^{-2}\rm s^{-1}$, equivalent to $\sim 4.9\%$ of the Crab Nebula flux
\citep{hegra_crab:1} above this threshold.

For each darkness period for which the significance of the observed
$\gamma$-ray excess exceeded $6\sigma$ (February, March, and April) a
differential spectrum was derived (see Fig.~\ref{fig_spectra}) and the
results of the corresponding power law fit according to
Eq.\,(\ref{powerlaw}) are listed in Table~\ref{table_spectrum}. Within
statistical errors, there is no indication of time variability of the
photon index. However, the changing flux normalisation $F_0$ indicates
source flux variability (see next section). Systematic errors were
estimated to be $\delta \Gamma \approx 0.2$ and $\delta F/F\approx
30\%$, dominated by the precision of the energy calibration of the
instrument and variations in the atmospheric extinction of the
Cherenkov light.

\begin{table}[h]
\caption{\label{table_spectrum}
Parameters of the power law fit to the differential spectrum for the
different darkness periods of H.E.S.S.\ observations on
PSR\,B1259$-$63. For the periods May and June 2004, no spectrum could
be derived due to insufficient statistics. Shown are the photon index
$\Gamma$ and the flux normalisation $F_\mathrm{0}$ (with statistical
error only), the $\chi^2$ per number of degrees of freedom
$\chi^2/\mathrm{ndf}$ and the $\chi^2$ probability $P_{\chi^2}$ for
the power law fit of the spectrum, and the corresponding mean
threshold energy $E_\mathrm{th}$ (rounded to 10\,GeV).}
\centering
\resizebox{\hsize}{!}{
\begin{tabular}{cccccc}
\hline\hline
Period
 & $\Gamma$
 & $F_\mathrm{0}$
 & $\chi^2/$
 & $P_{\chi^2}$
 & $E_\mathrm{th}$
\\
2004
 &
 & $\mathrm{[TeV}^{-1}\,\mathrm{cm}^{-2}\,\mathrm{s}^{-1}]$
 & $\mathrm{ndf}$
 &
 & $[\mathrm{GeV}]$
\\
\hline
February
 & $2.8\pm0.3$ & $2.1\pm0.3\times10^{-12}$ & $1.8/4$ & $0.77$ & 370\\
March
 & $3.0\pm0.6$ & $1.0\pm0.2\times10^{-12}$ & $1.8/4$ & $0.76$ & 420\\
April
 & $2.8\pm0.4$ & $2.0\pm0.3\times10^{-12}$ & $0.4/3$ & $0.93$ & 350\\
\hline
Overall$^*$
 & $2.7\pm0.2$ & $1.3\pm0.1\times10^{-12}$ & $2.3/5$ & $0.81$ & 380\\
\hline
\multicolumn{6}{l}{{\footnotesize $^*$ includes all data from February to May 2004}}\\
\end{tabular}
}
\end{table}

\subsection{Flux Variability}

\begin{figure*}[t]
\centering
\resizebox{\hsize}{!}{\includegraphics{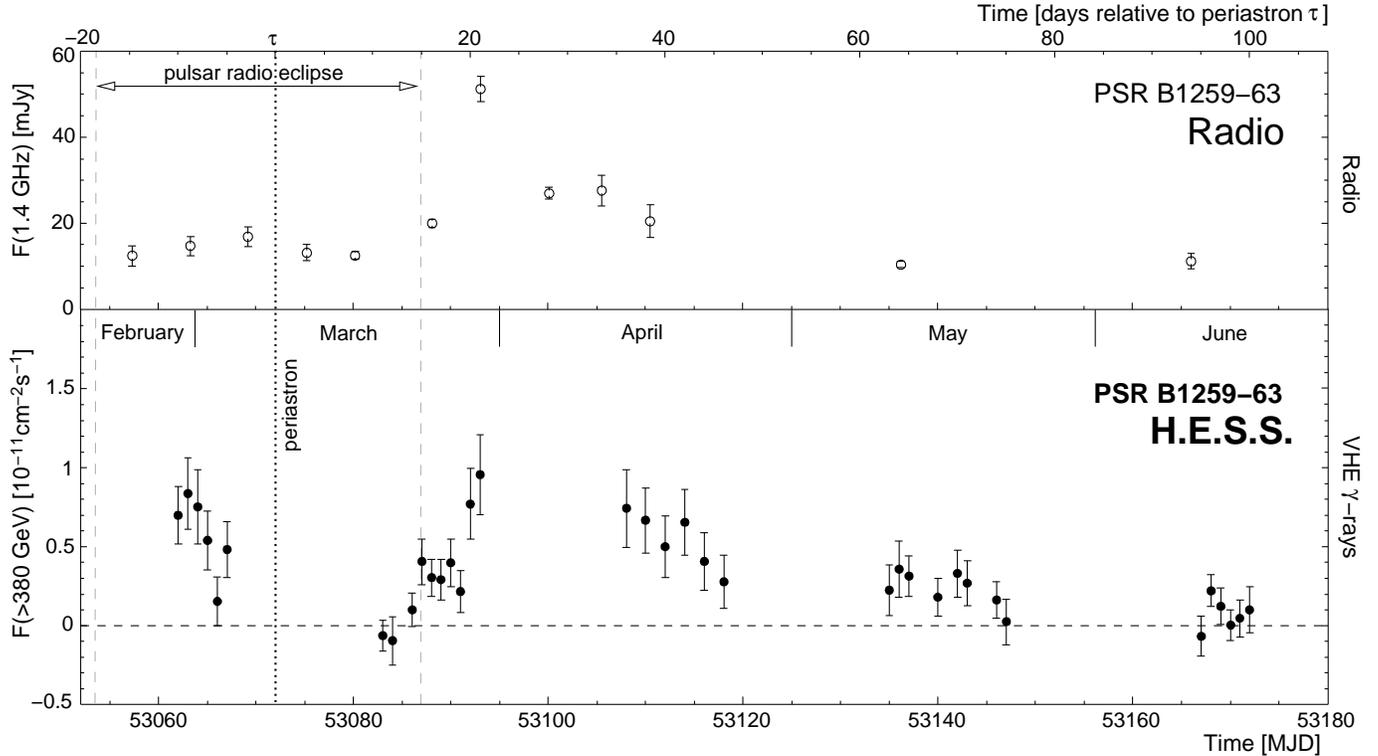}}
\caption{
VHE $\gamma$-ray and radio light curves of PSR\,B1259$-$63 around its
periastron passage (epoch $\tau$$=$0, dotted vertical line). {\em
Upper panel, open points}: Flux density of the transient unpulsed
radio emission at $1.4$\,GHz \citep{Johnston:5}. The pulsed radio
emission was eclipsed in the time interval between $\sim\tau-18$\,days
and $\sim\tau+15$\,days, indicated by the two dashed vertical
lines. {\em Lower panel, full points}: Daily integral flux above
380\,GeV as measured by H.E.S.S. Note that the moonlight prohibits
H.E.S.S.\ observations for $\sim10$\,days between each darkness
period.}
\label{fig_lightcurve}
\end{figure*}

The daily integral flux of $\gamma$-rays above the mean threshold of
380\,GeV is shown in Fig.~\ref{fig_lightcurve} (lower panel). Since
the spectrum cannot be derived on a daily basis due to limited
statistics, the integral flux $F(E>E_{\rm th}=380\,\rm GeV)$ was
obtained by integrating Eq.\,(\ref{powerlaw}) assuming the spectral
index of the time-averaged spectrum $\Gamma=2.7$. The flux
normalisation $F_0$ was calculated using the measured excess of
$\gamma$-rays $N_\gamma$ given by
\[
N_\gamma = F_0
\int{\mathrm{d}\Theta\,t_{\rm live}(\Theta)}
\int{\mathrm{d}E\,A_\mathrm{eff}(E,\Theta)\,\left(
\frac{E}{\mathrm{1\,TeV}}\right)^{-\Gamma}},
\]
with the daily live time $t_{\rm live}(\Theta)$ as a function of the
zenith angle $\Theta$, and the effective area $A_{\rm eff}(E,\Theta)$
after selection cuts obtained from simulations. Note that this method
does not depend on the energy reconstruction and so the complete
sample of excess events can be used. An alternative method, using the
energy reconstruction to calculate the flux based on the effective
area for each event above the energy threshold $E_{\rm th}$ according
to
\[
F(>E_{\rm th}) = \frac{1}{t_{\rm live}} \left\{
\sum^{E^\prime_i>E_{\rm th}}_{i,\rm on}
{\frac{1}{A_{\rm eff}(E^\prime_i,\Theta_i)}}
 - \alpha \sum^{E^\prime_j>E_{\rm th}}_{j,\rm off}
{\frac{1}{A_{\rm eff}(E^\prime_j,\Theta_j)}} \right\},
\]
with $E^\prime$ as the reconstructed energy and $\alpha$ as the
background normalisation, yields consistent results but introduces a
bigger statistical error since only events above the threshold are
considered.

The daily light curve in Fig.~\ref{fig_lightcurve} clearly indicates a
variable flux. This can be quantified by a fit of a constant flux to
the data yielding a $\chi^2$ of 90.9 per 35 degrees of freedom
corresponding to a $\chi^2$ probability of $1.2\times 10^{-7}$.

In order to investigate the flux trend in each darkness period, the
individual light curves were fitted separately by a straight line and
the significance for an increasing or decreasing flux is listed in
Table~\ref{table_lc}. The low flux state after periastron is followed
by a distinct rise beginning at $\tau\sim+15$\,days and a slow
decrease until $\tau\sim+75$\,days where the observed excess is no
longer significant.

\begin{table}[h]
\caption{\label{table_lc}
Light curve properties of the H.E.S.S.\ 2004 data. The integrated flux
$F(>380\,\mathrm{GeV})$ was calculated assuming a photon index $\Gamma
= 2.7$ (see text). For the daily integral flux light curve of each
period the following properties are given: the $\chi^2/\mathrm{ndf}$
for a fit to a constant flux, and the slope $a$ and significance $S_a
= \left|a\right|/\sigma_a$ of an increasing or declining flux after
fitting a straight line $F(t_{\rm MJD})=a t_{\rm MJD} + F(0)$ ($t_{\rm
MJD}$ is the Modified Julian Date), together with the corresponding
$\chi^2_{\rm line}/\mathrm{ndf}$.}
\centering
\resizebox{\hsize}{!}{
\begin{tabular}{cccccc}
\hline\hline
Period
 & $F(>380\,\mathrm{GeV})$
 & $\chi^2_\mathrm{const}/$
 & $a\;[10^{-9}\rm\,cm^{-2}$
 & $\chi^2_\mathrm{line}/$
 & $S_a$
\\
2004
 & $[10^{-12}\,\mathrm{cm}^{-2}\,\mathrm{s}^{-1}]$
 & $\mathrm{ndf}$
 & $\rm\,s^{-1}\,MJD^{-1}]$
 & $\mathrm{ndf}$
 & $[\sigma]$
\\
\hline
February
 & $5.9\pm0.6$ & $9.4/5$ & $-10\pm5$ & $4.3/4$ & 2.2 \\
March
 & $2.8\pm0.4$ & $31.9/9$ & $7.1\pm1.5$ & $8.9/8$ & 4.6 \\
April
 & $5.1\pm0.7$ & $4.2/5$ & $-4.4\pm2.4$ & $0.8/4$ & 1.8 \\
May
 & $2.4\pm0.5$ & $4.0/7$ & $-1.6\pm1.2$ & $2.2/6$ & 1.3 \\
June
 & $<2.0\,^*$ & $4.3/5$ & $-0.5\pm3.1$ & $4.3/4$ & 0.2 \\
\hline
\multicolumn{5}{l}{{\footnotesize $^*$ 99\% CL, calculated according to \citet{Feldman:1}}}\\
\end{tabular}
}
\end{table}

The upper panel of Fig.~\ref{fig_lightcurve} shows the light curve of
the transient unpulsed radio emission obtained from 2004 observations
of PSR\,B1259$-$63 \citep{Johnston:5} for the same time range as for
the TeV band. There is some correlation visible between the radio and
the TeV bands, especially a low flux state around periastron and a
high flux state at $\sim\tau+20$\,days. The pulsed radio emission from
the pulsar was eclipsed during the periastron passage
\citep{Johnston:1}, interpreted as due to scattering in the dense disk
material and allowing the dates of the disk crossings to be estimated,
in particular $\sim\tau-16$\,days and $\sim\tau+13$\,days for the 2004
periastron passage \citep{Johnston:5}. The observed dates of enhanced
VHE $\gamma$-ray flux roughly correspond to orbital phases a few days
after the assumed disk crossings. However, the coverage of the
considered time interval by H.E.S.S.\ observations is limited and does
not allow firm conclusions about VHE $\gamma$-ray flux maxima,
especially pre-periastron.

\section{Discussion and Conclusions}

The detection of VHE $\gamma$-rays from PSR\,B1259$-$63 gives the
first unambiguous and model-independent evidence of particle
acceleration to multi-TeV energies in this unique binary system. The
(most likely) synchrotron origin of X-rays \citep[e.g.][]{Tavani:2}
also indicates the presence of ultra-relativistic particles, in the
form of electrons, but other possible explanations of the X-radiation,
e.g.\ due to the bulk motion Comptonisation of the stellar photons by
the pulsar wind with a moderate Lorentz factor of 10--100
\citep{Chernyakova:1}, cannot be firmly excluded.

In contrast to the X-ray emission, the extension of the $\gamma$-ray
spectrum to several TeV necessarily implies that the parent particles,
electrons and/or protons, are accelerated to at least 10\,TeV. The
point-like feature of the detected signal constrains the extension of
the $\gamma$-ray production region to $R \leq 0.24\,\rm pc =
7.4\times10^{17}\,cm$. The information contained in the time
variability of the signal on timescales of days provides a hundred
times more stringent upper limit on the size of the $\gamma$-ray
source of about $10^{16}\,\rm cm$.

The most likely scenario of particle acceleration and radiation in
this system is a variety of the standard model of Pulsar Driven
Nebulae \citep[see e.g.][]{Rees:1,Kennel:1} which postulates that the
deceleration of the ultra-relativistic pulsar wind (with a bulk motion
Lorentz factor of $10^6$--$10^7$) by the pressure forces of the
external medium enforces a termination shock. In this context
electrons are accelerated to multi-TeV energies. These electrons
radiate in the magnetic and photon fields in which they propagate and
thus produce synchrotron and Inverse-Compton nebulae with a typical
size exceeding 0.1\,pc, commonly called {\em plerions}.

The mean energy flux contained in the VHE $\gamma$-ray emission
derived from the time-averaged energy spectrum is $EF_E \approx 3
\times 10^{-12}\rm\,erg\,cm^{-2}\,s^{-1}$ which represents a $\gamma$-ray
luminosity of $L_\gamma \approx 8 \times 10^{32}\rm\,erg\,s^{-1}$ for
an assumed distance of the binary system of $1.5\,$kpc. This
corresponds to $\approx0.1\%$ of the pulsar spin-down
luminosity. Interestingly, this value matches the typical ratio
between the pulsed and unpulsed component of X-ray radiation from
isolated pulsars with associated pulsar wind nebulae and their
spin-down luminosity \citep[see e.g.\ ][]{Cheng:1}, supporting the
suggestion that the energy for the observed $\gamma$-rays is provided
by the pulsar.

However, in the case of PSR\,B1259$-$63 both the spatial and temporal
scales are expected to be rather different from those in plerions
around isolated pulsars. Also, dynamical flow effects can play a
different role since the pulsar moves through the stellar wind and the
photon field of the massive companion star. Due to the high pressure
of the stellar wind, the pulsar wind terminates very close to the
pulsar and thus the electrons are accelerated well within the binary
system. Moreover, because of severe adiabatic and radiative losses,
the TeV electrons have a very short lifetime. As a result, the
radiation is emitted in a rather compact region not far from the
acceleration site, and for any given time the emission originates from
a quite short sector of the pulsar trajectory.

In the following we summarise the physical processes expected to be at
work, and put our observational results into this perspective.

\subsection{Characteristic Cooling Times}

The synchrotron loss time
\begin{equation}
t_{\rm sy}=6 \pi \frac{m_{\rm e}^2 c^4}{\sigma_{\rm T} c \epsilon B^2}
\simeq 400\,\epsilon_{\rm TeV}^{-1} B_{\rm G}^{-2}\,\rm s\,,
\label{t_synch}
\end{equation}
of an electron with energy $\epsilon$ depends on the local magnetic
field strength $B$~(where $B_G = B / 1\,G$) which is not known {\em a
priori}. We shall estimate $B$ below by combining the {\em INTEGRAL}
X-ray data and our H.E.S.S.\ TeV $\gamma$-ray data. In contrast, we
can estimate the inverse Compton cooling rate of electrons with very
good accuracy at any given time, i.e.\ at any location of the pulsar
in its orbit \citep{Tavani:2,Kirk:1}.

Indeed, the inverse Compton losses are dominated by the scattering of
electrons on the light of the companion star with luminosity $L_{\rm
star}=3.3 \times 10^{37}\,\rm erg/s$ and effective blackbody
temperature $T_{\rm eff}=2.3 \times 10^4\,\rm K$. At any distance $R$,
the black-body radiation decreases by a factor of $(R/R_\star)^{-2}$,
where $R_\star=4 \times 10^{11}\,\rm cm$ is the radius of the
companion star. Thus at the distance of the pulsar to the star
$R_{13}=R/10^{13}\,\rm cm$, the energy density of the radiation is
\begin{equation}
w_{\rm r}= \frac{L_{\rm star}}{4 \pi R^2 c} \simeq 0.87
R_{13}^{-2}\,\rm erg/cm^3\,,
\label{w_r}
\end{equation}
and correspondingly the electron cooling time is determined in the
deep relativistic Klein-Nishina regime which applies here for TeV
energies, given the high energies of the seed photons. For a Planckian
seed photon spectrum, and using the formalism of \citet{Blumenthal},
one obtains
\begin{equation} 
t_{\rm KN}=3.2 \times 10^3 \frac{\epsilon_{\rm TeV} R_{13}^2}{\ln
(30.6 \epsilon_{\rm TeV}) -1.4}\,\rm s\,.
\label{t_KN}
\end{equation} 

\begin{figure}[t]
\centering
\resizebox{\hsize}{!}{\includegraphics{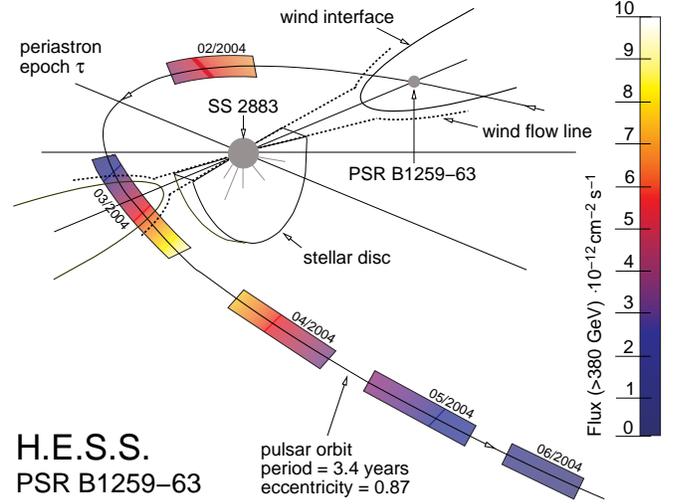}}
\caption{
Sketch of the orbit of PSR\,B1259$-$63 with respect to the line of
sight \citep[adapted from][]{Johnston:3}. The pulsar approaches the
equatorial disk prior to periastron while it is ``behind'' the
companion star and turns towards the observer before it crosses the
disk after periastron for the second time. Note that the orientation
of the disk with respect to the orbital plane is not precisely known.
The colour gradient bars along the orbit indicate the periods of
H.E.S.S.\ observations and show the integral VHE $\gamma$-ray flux
using a smoothed light curve based on the data points from
Fig.~\ref{fig_lightcurve}.}
\label{fig_cartoon}
\end{figure}

For PSR\,B1259$-$63, with its environment of a strong stellar wind and
a strong radiative flow from the companion, we must in addition
consider the adiabatic losses of the accelerated pulsar
particles. This has already been noted by \citet{Tavani:2}. Indeed,
the plasma flow beyond the pulsar wind termination shock differs
strongly from the spherically symmetric configuration of an isolated
pulsar in a static pressure environment. Rather than being decelerated
in a spherically symmetric fashion while compressing the post-shock
magnetic field, the actual flow will stagnate at the sub-stellar point
at a distance $l\sim 10^{12}$\,cm and expand relativistically into the
downstream direction, which roughly points away from the Be star. The
surface of separation between the interior flow of the pulsar
particles and the exterior stellar wind flow is indicated as ``Wind
interface'' in Fig.~\ref{fig_cartoon}.

The expansion of the relativistic, shocked pulsar wind gas will
proceed with a flow velocity of magnitude $|\underline{u}|$, roughly
equal to the relativistic magnetosonic speed $\varepsilon v_{\rm
ms,\perp}$, where $v_{\rm ms,\perp}\simeq c \sqrt{2/3}$, over a
distance $\Delta l$, where $\Delta l$ is a few $\times l$. The factor
$\varepsilon$ ranges between $\sqrt{2}/2$ and 1, if the direction of
the postshock MHD flow ranges between a perpendicular and a parallel
orientation with respect to the magnetic field direction,
respectively. Assuming the pitch angle scattering mean free path of
the accelerated pulsar wind particles to be small compared to $l$,
their adiabatic loss time is then given by
\begin{equation}
t_{\rm ad} = \frac{3 \varepsilon}{\mathrm{div}\,\underline{u}}
\simeq \varepsilon \times 350
\frac{\Delta l}{2\times 10^{12}\,\mathrm{cm}}\,\rm s\,,
\end{equation}
with $0.71 < \varepsilon < 1$, depending on the details of the post-shock
MHD flow.

In this picture the adiabatic losses proceed about as fast as the
synchrotron losses if $B \approx 1$\,G (compare with
Eq.\,(\ref{t_synch})) -- while both are faster than the inverse Compton
losses -- and the energy spectrum of the radiating electrons is close
to their source spectrum produced during acceleration at the
termination shock.
We shall consider neither electron Bremsstrahlung nor $\gamma$-ray
production in interactions of relativistic protons with the ambient
gas via decay of secondary $\pi^0$-mesons
\citep[see~e.g.][]{Kawachi:1} here, since the accelerated pulsar wind
particles are hydromagnetically constrained to the magnetic field
lines of the pulsar outflow and will hardly interact individually with
the stellar wind particles, except in a thin boundary layer at the
Wind interface far downstream.

The missing quantity in our cooling considerations is the value of
$B$. We shall use the contemporaneous {\em INTEGRAL} results and our
H.E.S.S.\ fluxes to estimate it as follows: Since there is little
doubt that the detected TeV $\gamma$-rays are produced in the
Klein-Nishina regime, we may conclude that they are emitted by
electrons of the same energy, $\epsilon_e \approx E_\gamma$. This
allows to set up a direct relation between $\gamma$-rays and X-rays
produced by the same electrons,
\begin{equation}\label{X_gamma}
E_{\rm X} > 20 B_{\rm G} E_{\rm TeV}^2\,\rm keV\,.
\end{equation}
The Compton losses with $E_{\rm TeV}$ take place in a spatial region
that is larger than that of synchrotron losses, and therefore
adiabatic losses have already occurred there.

An experimental estimate of $B$ is possible by comparing the energy
fluxes of the detected X- and $\gamma$-rays. The post-periastron data
obtained in March by {\em RXTE} and {\em INTEGRAL} in the X-ray band
from 1\,keV to 100\,keV \citep[see][]{Shaw:1} show a relatively flat
spectral energy distribution $\nu F_\nu \sim E^2 {\rm d}N/{\rm d}E$ at
the level of $3\times 10^{-11}\,\rm erg/c m^2\,s$. The VHE
$\gamma$-ray flux for the same period was $\approx 2 \times
10^{-12}\,\rm erg/s$. This implies that in the TeV energy regime the
synchrotron losses of electrons proceed an order of magnitude faster
than the inverse Compton losses. Therefore, from the comparison of the
synchrotron cooling time, given by Eq.\,(\ref{t_synch}), with the
Klein-Nishina cooling time given by Eq.\,(\ref{t_KN}), and taking into
account that two weeks after the periastron the separation between the
pulsar and the companion star was approximately $1.5 \times
10^{13}\,\rm cm$, we find

\begin{equation}
B \approx 1 E_{\rm TeV}^{-1}\,\rm G.
\label{B_est}
\end{equation}

Our experimentally determined fields of the order of 1\,G agree rather
well with the theoretical estimate of the field derived from the MHD
treatment of the pulsar wind \citep{Tavani:2,Ball:3}, and the value
deduced from the observed unpulsed radio emission near periastron
\citep{Connors:1}.

\begin{figure}[th]
\centering
\resizebox{\hsize}{!}{\includegraphics{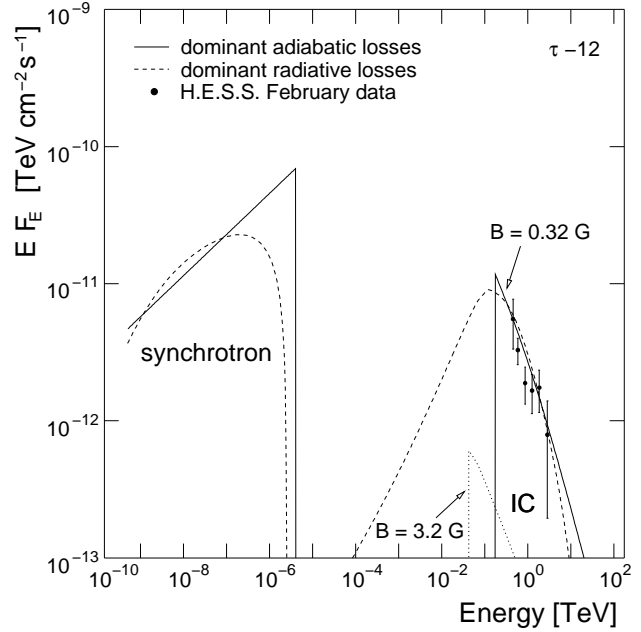}}
\caption{\label{fig_sed}
Spectral energy distribution from \citet{Kirk:1} assuming the electron
cooling to be dominated by radiative (dashed lines) or adiabatic
losses (solid lines) for a magnetic field strength of 0.32\,G at an
orbital phase 12 days prior to periastron. The spectra were calculated
by \citet{Kirk:1} for different electron injection spectra with the
power law indices $\alpha_0=1.4$ and $\alpha_0=2.4$, respectively. For
illustration, the predicted spectrum for adiabatic losses for
$B=3.2\,$G is also shown (dotted line). Data points represent the
measured energy flux from H.E.S.S.\ observations in February 2004.}
\end{figure}

\subsection{Spectral and Temporal Characteristics of the TeV Radiation}
\label{lc_discussion}

Let us assume that the accelerated electrons enter the $\gamma$-ray
production region with a rate $Q\,(\epsilon) \propto
\epsilon^{-\alpha_0}$. As a consequence, in the case of dominant
synchrotron or inverse Compton losses in the Thompson regime, a
steeper electron spectrum is established in the $\gamma$-ray
production region with power-law index $\alpha=\alpha_0+1$, while
dominating adiabatic (or Bremsstrahlung losses) will not change the
initial electron spectrum, i.e.\ $\alpha=\alpha_0$, both for
$\alpha_0>1$.

Let us conversely assume that the radiating electrons have a power law
distribution ${\rm d} N/\rm d \epsilon \propto
\epsilon^{-\alpha}$. Then the inverse Compton scattering in the
Klein-Nishina regime leads to a $\gamma$-ray spectrum
\citep{Blumenthal}
\begin{equation}\label{IC_spec}
\frac{{\rm d} N}{{\rm d}E} \propto E^{-(\alpha+1)} \left(\ln\frac{k_{\rm B}T\,E}{m_{\rm e}^2 c^4} + C\right).
\end{equation}
Using the H.E.S.S.\ results, the photon index of the radiating
electrons can therefore be estimated from the observed $\gamma$-ray
spectrum ${\rm d} N/{\rm d}E \propto E^{-\Gamma_\gamma}$ with
$\Gamma_{\gamma} = 2.7 \pm 0.2_{\rm stat}$ by equating it with the
expected Klein-Nishina spectrum of Eq.~(\ref{IC_spec}). A more
detailed numerical evaluation of the IC $\gamma$-ray spectrum shows
that in our energy range the average photon spectral index can be
approximated as $\Gamma_\gamma \approx \alpha+0.5$. Therefore we find
$\alpha \approx 2.2$. In the scenario in which adiabatic losses
dominate, this is, within errors, compatible with an acceleration
process giving $\alpha_0=2$. If radiation losses dominate, this
requires an injection spectrum significantly harder than $\alpha_0=2$.

\begin{figure*}[th]
\centering
\resizebox{\hsize}{!}{\includegraphics{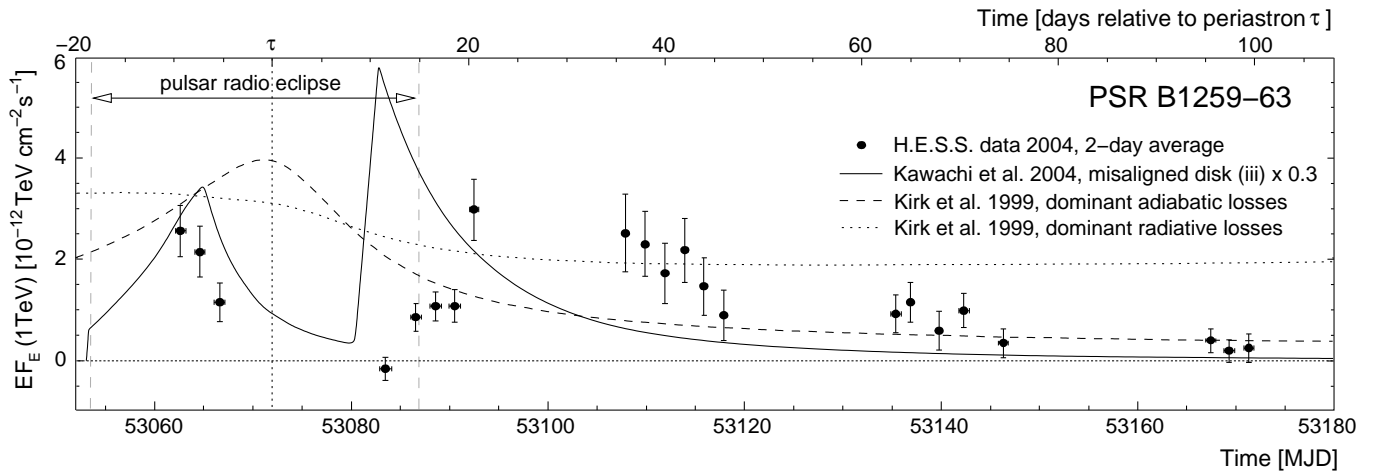}}
\caption{
Comparison of the light curves predicted by different models of VHE
$\gamma$-ray emission with H.E.S.S.\ data. The solid line was taken
from \citet{Kawachi:1}, for a misaligned disk geometry, model (iii) in
Fig.~7, and scaled by a factor of 0.3. The dashed and dotted lines
represents the curves given in \citet{Kirk:1}, for dominant adiabatic
and radiative energy losses, respectively. The full points represent
the corresponding light curve obtained from the H.E.S.S.\ data using a
bin width of 2\,days.}
\label{fig_lc_models}
\end{figure*}

The relevant photon spectral index range $1.6 < \Gamma_X < 1.8$ in the
$1 \leq E_X \leq 10$\,keV energy range from the 1994 {\em ASCA}
observations \citep{Hirayama:1}, although not coeval, would also
indicate a rather hard spectrum of the radiating electrons $2.2 \leq
\alpha \leq 2.6$ which would again be difficult to reconcile with
dominant synchrotron cooling. This 1994 synchrotron photon spectral
index is consistent with a preliminary spectral fit with $\Gamma_X =
1.82 \pm 0.34$ derived from the coeval {\em INTEGRAL} observations
\citep{Shaw:2}. In this paper we shall ultimately rely on the
H.E.S.S.\ spectrum.

Within the scheme of dominant adiabatic cooling, the strong
variability of the VHE $\gamma$-ray emission might be explainable by a
varying spatial confinement of the accelerated pulsar wind particles
by the kinetic and thermal pressure of the stellar mass outflow. In
Figure~\ref{fig_cartoon} the variations of the measured VHE
$\gamma$-ray flux are illustrated within the context of the orbital
parameters and environment of the binary system with respect to the
line of sight. The enhanced mass outflow in the equatorial disk will
result in a more compact emission region -- and therefore weaker
adiabatic cooling -- during the pulsar disk passages compared to the
other regions of the orbit, in particular the periastron phase, where
the expected stronger adiabatic cooling could lead to a minimum of the
$\gamma$-ray flux. A second flow aspect would be a Doppler modulation
due to the relativistic bulk flow velocity away from the stagnation
point of the shocked pulsar wind. Considering Fig.~\ref{fig_cartoon},
this could again lead to a flux minimum during periastron. However, a
detailed consideration of this overall physical picture is beyond the
scope of the present paper.

A minimum in the X-ray flux, observed for previous periastron
passages, has been interpreted as being due to increased IC losses
\citep{Tavani:2}. However, H.E.S.S.\ measurements indicate a minimum
in the IC flux around periastron, which would rule out such an
interpretation for this periastron passage. Unfortunately, due to the
full moon and bad weather conditions we were not able to take data
during the periastron passage of the pulsar.

The predicted pre-periastron spectrum from \citet{Kirk:1} is compared
to the corresponding H.E.S.S.\ data in Fig.~\ref{fig_sed}. The model
curves, taken from \citet{Kirk:1} (Figs.~2,6), were computed for the
two scenarios of dominant radiative losses and dominant adiabatic
losses, and matched the archival X-ray data for electron injection
spectra with $\alpha_0=1.4$ and $\alpha_0=2.4$, respectively. They
agree well with the TeV data, assuming a magnetic field strength of
$B=0.32\,$G supporting the estimate of the order of 1\,G obtained from
the H.E.S.S.\ data. Despite the spectral agreement, the models fail to
describe the TeV light curve as can be seen in
Fig.~\ref{fig_lc_models} (dashed and dotted lines). In particular,
neither the minimum around nor the high flux states before and after
periastron are described by the model. This is perhaps not surprising,
because the model assumed that the ratio between the energy densities
of the magnetic field and the photon field does not change throughout
the full orbit, and that the effects related to the pulsar passage
through the stellar disk can be neglected.

Interestingly, the observed light curve seems to be qualitatively
similar to the prediction made by the model of
\citet{Kawachi:1} which is also shown in Fig.~\ref{fig_lc_models}
(solid line). In this model, hadronic interactions and $\pi^0$
production in the misaligned stellar disk plays a dominant role in the
$\gamma$-ray production mechanism. However, present data do not allow
safe conclusions concerning the interpretation of the TeV light
curve. Studies of the most interesting part of the pulsar orbit have
to be postponed until multi-wavelength observations during periastron
become possible.

In summary, the detection of VHE $\gamma$-ray emission from the binary
pulsar PSR\,B1259$-$63 by H.E.S.S.\ provides the first
model-independent evidence of particle acceleration in this
object. The results clearly demonstrate the power of $\gamma$-ray
observations for the study of the properties and the nature of high
energy processes in this unique cosmic accelerator. Further
observations at VHE $\gamma$-ray energies are necessary to derive
spectra on a daily basis also during periastron passage.

\begin{acknowledgements}

The support of the Namibian authorities and of the University of
Namibia in facilitating the construction and operation of H.E.S.S.\ is
gratefully acknowledged, as is the support by the German Ministry for
Education and Research (BMBF), the Max Planck Society, the French
Ministry for Research, the CNRS-IN2P3 and the Astroparticle
Interdisciplinary Programme of the CNRS, the U.K. Particle Physics and
Astronomy Research Council (PPARC), the IPNP of the Charles
University, the South African Department of Science and Technology and
National Research Foundation, and by the University of Namibia. We
appreciate the excellent work of the technical support staff in
Berlin, Durham, Hamburg, Heidelberg, Palaiseau, Paris, Saclay, and in
Namibia in the construction and operation of the equipment. We also
thank S.~Shaw for informations on the photon spectral index from the
{\it INTEGRAL} X-ray observations.

\end{acknowledgements}

\bibliographystyle{aa}
\bibliography{psrb1259_aa}

\end{document}